%% file: main.tex
\documentclass[conference]{IEEEtran}
\IEEEoverridecommandlockouts

\usepackage[utf8]{inputenc}
\usepackage{blindtext}
\usepackage{color}
\usepackage{filecontents}

\usepackage{bbm}
\usepackage{array}
\usepackage{url}
\makeatletter
\@ifclasslater{IEEEtran}{2012/12/26}
 {\newcommand{\killproofname}{\unskip\nopunct}}
 {\newcommand{\killproofname}[1]{\unskip\aftergroup\ignorespaces\ignorespaces}}
\makeatother

\usepackage{graphicx}
\usepackage{subcaption}
\usepackage{subfiles}
\usepackage{amsfonts}
\usepackage{textcomp}
\usepackage{gensymb}
\usepackage{hyperref}
\usepackage{amsthm,amsmath}
\usepackage{cleveref}

\usepackage{svg, import}

\usepackage{bm}
\let\labelindent\relax
\usepackage{enumitem}

\newtheorem{thm}{Theorem}
\newtheorem{lemma}[thm]{Lemma}

\newtheorem{corollary}[thm]{Corollary}

\theoremstyle{definition}
\newtheorem{definition}{Definition}

\makeatletter
\@addtoreset{proofpart}{thm}
\makeatother

\newcommand{\ind}[1]{^{\left(#1\right)}}
\newcommand{\vect}[1]{\mathbf{#1}}
\newcommand{\mat}[1]{\mathbf{#1}}
\newcommand{\spikes}[1]{\left\lbrace t_\ell\ind{#1}, \ell = 1\cdots n^{(#1)}_{\mathrm{spikes}} \right\rbrace}

\newcommand{\bias}{\beta}
\newcommand{\biasi}[1]{\beta\ind{i}}

\newcounter{assumptCounter}

\newenvironment{assumptions}{
    \begin{enumerate}[label = (A\arabic*)]
    \setcounter{enumi}{\value{assumptCounter}}
    }
    {
    \setcounter{assumptCounter}{\value{enumi}}
    \end{enumerate}
    }

\usepackage{todonotes}
\usepackage{tikz}
\usetikzlibrary{math}
\graphicspath{{figures/}{../figures/}}

\begin{document}

\setcounter{assumptCounter}{0}
\title{How Asynchronous Events Encode Video}

\def\FUND{This work was supported by the Swiss National Science Foundation grant number 200021\_181978/1, ``SESAM - Sensing and Sampling: Theory and Algorithms''.}

    \author{\IEEEauthorblockN{Karen Adam, Adam Scholefield, and Martin Vetterli\thanks{\FUND}}
\IEEEauthorblockA{\textit{School of Computer and Communication Sciences, Ecole Polytechnique F\'ed\'erale de Lausanne, Lausanne, Switzerland} \\
firstname.lastname@epfl.ch}}

\maketitle
\subfile{sections/0_abstract.tex}
\subfile{sections/1_introduction.tex}

\subfile{sections/2_background.tex}

\subfile{sections/6_3DSignals.tex}

\subfile{sections/Conclusion.tex}

\IEEEpeerreviewmaketitle

\bibliographystyle{IEEEtran}
\bibliography{main}

\end{document}

%% file: sections/0_abstract.tex
\begin{abstract}
As event-based sensing gains in popularity, theoretical understanding is needed to harness this technology's potential.
Instead of recording video by capturing frames, event-based cameras have sensors that emit events when their inputs change, thus encoding information in the timing of events. This creates new challenges in establishing reconstruction guarantees and algorithms, but also provides advantages over frame-based video.
We use time encoding machines to model event-based sensors: TEMs also encode their inputs by emitting events characterized by their timing and reconstruction from time encodings is well understood. We consider the case of time encoding bandlimited video and demonstrate a dependence between spatial sensor density and overall spatial and temporal resolution.
Such a dependence does not occur in frame-based video, where temporal resolution depends solely on the frame rate of the video and spatial resolution depends solely on the pixel grid. However, this dependence arises naturally in event-based video and allows oversampling in space to provide better time resolution. As such, event-based vision encourages using more sensors that emit fewer events over time.

\end{abstract}


%% file: sections/1_introduction.tex
\section{Introduction}

The current approach to recording video---where multiple pictures are taken in close succession to each other--- evolved as it did because techniques to take pictures were established first, and video was developed as an extension thereof.

While there is no doubt that this approach works very well, there is no guarantee that the approach is optimal. Assume, for example, that you are taking a video of a bird against a blue sky background. To record this video, every pixel records the intensity of the light it receives for \textit{every} frame. With a frame rate of e.g. 60 frames per second, given that the sky is large, and that the video is potentially longer than a fraction of a second, the recorded data  is inevitably redundant.

The above example illustrates why frame-based video is suboptimal from a sample-complexity perspective, i.e. many samples are needed to encode little information. There are also inconveniences from a hardware-implementation perspective. For example, recording such a video has high power requirements (in large part due to the need to quantize the output, requiring high signal-to-noise ratio), and such a recording has limited dynamic range (also due to necessary quantization).

\begin{figure*}[tb]
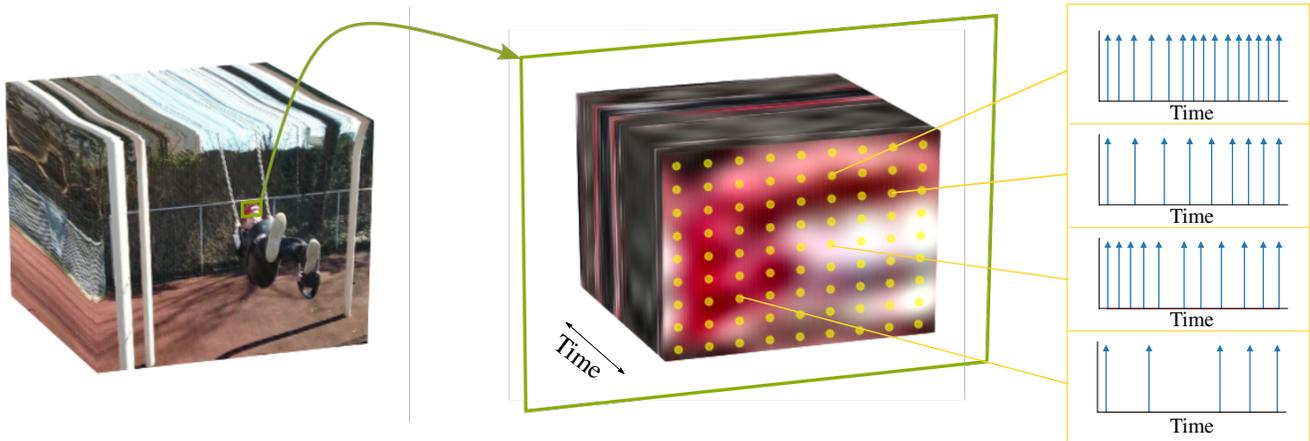

     \begin{minipage}[b]{\linewidth}
        \centering
            \def\svgwidth{1\columnwidth}%

                \subfile{../figures/tem_vid.tex}

     \end{minipage}
    \caption{Vision setup: we assume that we have an array of spiking devices, such as photoreceptors or TEMs, each of which is observing a scene at a particular location. The input to the receptor at this location is a time varying signal and the receptor will output a stream of spikes, the timing of which is dependent on the input.
    On the left, we show the projection of the scene which is being observed. In the middle, we show a patch of this scene, which is interpolated under bandlimited periodic assumptions, with an overlay of event-based sensors shown in yellow. To its right, we zoom in to view the spiking output of some of the sensors. The video used is taken from the Need for Speed dataset~\cite{galoogahi2017need}.}
    \label{fig: Figure 1}
    \end{figure*}

Fortunately, recent advances in event-based video can help counter these issues. Event-based cameras have pixels that each emit an event whenever their input exhibits a change that is ``large enough''~\cite{delbruck2010activity,gallego2019event}. The output of an event-based camera is a stream of events associated with each pixel, where the streams are different across pixels, and the timings of the events depend on the input to the pixel which emits it, as depicted in Fig.~\ref{fig: Figure 1}. In this encoding paradigm, frames are obsolete, and information is recorded when the sensors detect new information.

Consequently, event-based video has an improved sample complexity as it reduces redundancy in the recorded information: when a pixel's input does not change, this pixel emits no event. Moreover, as the timing of the events holds the information, events can be emitted with very low power, as long as they can be detected. In classical sampling, the need for amplitude quantization implies a need for a high signal-to-noise ratio at the output, thus requiring high power at the output. In event-based sensing, if quantization is needed, it is performed in the time dimension and does not require high power, but rather a fast clock and the ability to recognize an event at the output. The lack of amplitude quantization also resolves issues with limited dynamic range as one no longer needs quantization levels that are fixed ahead of time~\cite{rebecq2019high} and quantization in time can reach much more accurate levels.

However, as event-based vision is a new technology, it is less well understood than frame-based vision. One key difficulty with understanding event-based vision is the fact that different sensors emit events at different times, making it difficult to build frames and thus difficult to perform a reconstruction.

In this paper, we will show that the asynchrony of events across pixels actually consitutes an advantage of event-based vision over frame-based vision and allows an encoding of information that is more sample-efficient.

To do so, we use time encoding machines (TEMs) as a model for event sensors to understand how these sensors encode their data. TEMs are devices that integrate their inputs and emit events (or \emph{fire spikes}) when the integral reaches a threshold, and the information about the input is encoded in the timing of the events or spikes, much as in the case of event sensors. Moreover, the reconstruction properties of TEMs are well understood for different classes of inputs~\cite{lazar2004perfect,alexandru2019reconstructing,hilton2021guaranteed,rudresh2020time} and configurations~\cite{lazar2008faithful,gontier2014sampling}. We will see that using event sensors or TEMs results in streams of events that are asynchronous across sensors, thus allowing for richer information content than in the frame-based approach. To do so, we assume our scene can be modeled by a periodic bandlimited function in three dimensions (two spatial and one temporal dimension) and will see how the setup results in an entanglement between spatial and temporal resolution. In the frame based case, spatial and temporal resolution are uniquely and respectively defined by the pixel gridding and frame rate, but in the case of event-based vision, we will see that the temporal resolution is also affected by the pixel gridding. As a result, temporal resolution in event-based vision can be increased by increasing the number of pixels, thus encouraging a generally higher number of pixels in a camera and lower firing rate per pixel.

While results in this paper are covered in~\cite{adam2021asynchrony}, we here provide a thematic and self-contained approach to understanding event-based video from a time encoding perspective. More detailed theoretical results can be found in~\cite{adam2021asynchrony}.

%% file: figures/tem_vid.tex
\begingroup%
  \makeatletter%
  \providecommand\color[2][]{%
    \errmessage{(Inkscape) Color is used for the text in Inkscape, but the package 'color.sty' is not loaded}%
    \renewcommand\color[2][]{}%
  }%
  \providecommand\transparent[1]{%
    \errmessage{(Inkscape) Transparency is used (non-zero) for the text in Inkscape, but the package 'transparent.sty' is not loaded}%
    \renewcommand\transparent[1]{}%
  }%
  \providecommand\rotatebox[2]{#2}%
  \newcommand*\fsize{\dimexpr\f@size pt\relax}%
  \newcommand*\lineheight[1]{\fontsize{\fsize}{#1\fsize}\selectfont}%
  \ifx\svgwidth\undefined%
    \setlength{\unitlength}{722.38375842bp}%
    \ifx\svgscale\undefined%
      \relax%
    \else%
      \setlength{\unitlength}{\unitlength * \real{\svgscale}}%
    \fi%
  \else%
    \setlength{\unitlength}{\svgwidth}%
  \fi%
  \global\let\svgwidth\undefined%
  \global\let\svgscale\undefined%
  \makeatother%
  \begin{picture}(1,0.32966687)%
    \lineheight{1}%
    \setlength\tabcolsep{0pt}%
    \put(0,0){\includegraphics[width=\unitlength,page=1]{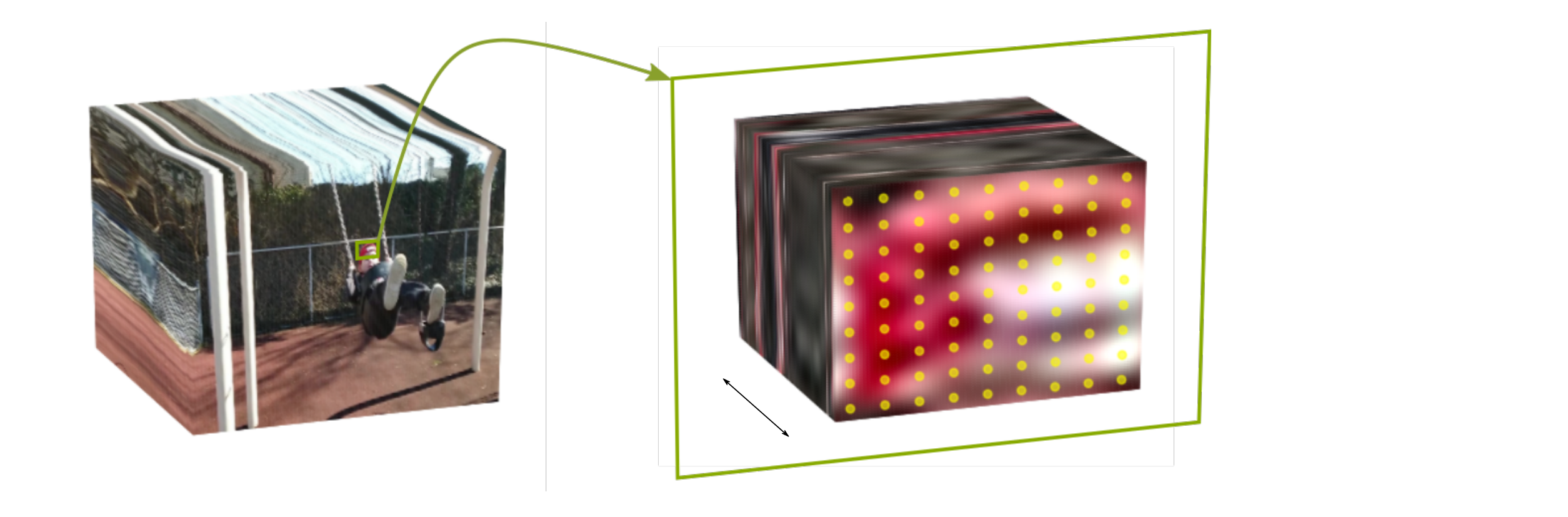}}%
    \put(0.45092433,0.07046197){\color[rgb]{0,0,0}\rotatebox{-37.210289}{\makebox(0,0)[lt]{\lineheight{1.25}\smash{\begin{tabular}[t]{l}Time\end{tabular}}}}}%
    \put(0,0){\includegraphics[width=\unitlength,page=2]{tem_vid_pdf.pdf}}%
    \put(0.89724911,0.00874166){\color[rgb]{0,0,0}\makebox(0,0)[lt]{\lineheight{1.25}\smash{\begin{tabular}[t]{l}\footnotesize Time\end{tabular}}}}%
    \put(0.89724911,0.23507591){\color[rgb]{0,0,0}\makebox(0,0)[lt]{\lineheight{1.25}\smash{\begin{tabular}[t]{l}\footnotesize Time\end{tabular}}}}%
    \put(0.89724911,0.16032352){\color[rgb]{0,0,0}\makebox(0,0)[lt]{\lineheight{1.25}\smash{\begin{tabular}[t]{l}\footnotesize Time\end{tabular}}}}%
    \put(0.89724911,0.08557084){\color[rgb]{0,0,0}\makebox(0,0)[lt]{\lineheight{1.25}\smash{\begin{tabular}[t]{l}\footnotesize Time\end{tabular}}}}%
    \put(0,0){\includegraphics[width=\unitlength,page=3]{tem_vid_pdf.pdf}}%
  \end{picture}%
\endgroup%

%% file: sections/2_background.tex
\section{Background}
\label{sec: background}

\begin{figure}[tb]
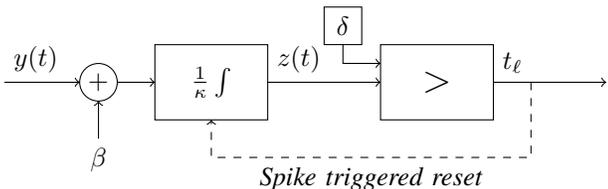

	\begin{minipage}[b]{0.85\linewidth}
		\centering
        \subfile{../figures/TEMCircuit_Tikz.tex}
	\end{minipage}
	\vspace{-2em}
	\caption{Circuit of a Time Encoding Machine, with input $y(t)$, threshold $\delta$, integrator constant $\kappa$ and bias $\bias$.}
	\label{fig:TEM circuit}
	\end{figure}

We use a time encoding machine (TEM) as a model for an event-based sensor~\cite{lazar2003time}, where the TEM follows an integrate and-fire mechanism and encodes its input using \emph{times} that are dependent on the input itself, depicted in Fig.~\ref{fig:TEM circuit}.

\begin{definition}
	\label{def: tem}
    A \textit{time encoding machine} (TEM) with parameters $\kappa$, $\delta$, and $\bias$ takes an input signal $y(t)$, adds a bias $\bias$ to it and integrates the result, scaled by $1/\kappa$, until a threshold $\delta$ is reached. Once this threshold is reached, the time $t_\ell$ at which it is reached is recorded, the value of the integrator resets to $-\delta$ and the mechanism restarts. We say that the machine \emph{spikes} at the integrator reset and call the recorded time $t_\ell$ a \textit{spike time}.
\end{definition}

We adopt the integrate-and-fire model for the event-based sensor because it is well understood and results on sampling and reconstruction are widely available. However, the results can be extended to different types of time encoding machines which filter the signal, compare it to an output and emit events when the comparison yields a match~\cite{gontier2014sampling}. A differentiate-and-fire model, for example, can be written similarly, as well as more complicated event-based sensing models.

It was shown that a bandlimited signal can be perfectly recovered from its time encoding as described in Definition~\ref{def: tem} if the parameters of the TEM
ensure a maximal spacing between consecutive spikes of a TEM, thus ensuring that a Nyquist-like condition on the measurements is satisfied.

Moreover, the reconstruction of the input uses the fact that the timing of the emitted spikes provides linear constraints on the input. In fact, each pair of consective spike times $t_\ell$ and $t_{\ell+1}$ provides the value of the integral of the input:
\begin{equation}
	\int_{t_\ell}^{t_{\ell+1}}y(u)\, du = 2\kappa\delta - \bias(t_{\ell+1}-t_\ell).
	\label{eq: sig integral}
	\end{equation}

In this paper, we tackle the problem of time encoding video, achieved by replacing pixels in a standard frame-based camera by event-based sensors or, in this case, time encoding machines, as depicted in Fig.~\ref{fig: Figure 1}. Video time encoding machines have been examined before, and results on perfect reconstruction have already been established~\cite{lazar2011video}. However, previous work relied on applying linearly independent filters to a bandlimited video before it is processed by the TEMs. Here, we propose a filter-less approach where the scene is processed without preprocessing. Moreover, we clarify a dependency between spatial sampling density and temporal resolution that was not apparent before.

    \begin{figure}[tb]
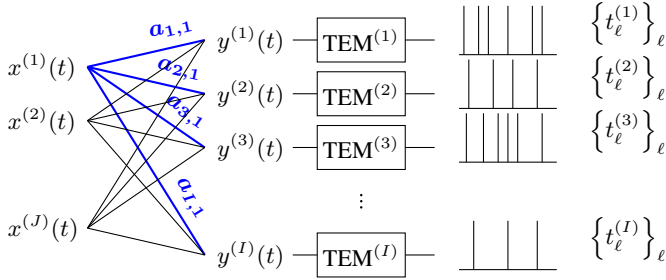

	\begin{minipage}[b]{0.85\columnwidth}
        \centering
        \subfile{../figures/Mixed_TEM.tex}
		\end{minipage}
    \caption{Sampling setup: $J$ input signals $x\ind{j}(t)$, $j=1\cdots J$ are mixed using a matrix $\mat{A}$ and produce signals $y\ind{i}(t)$, $i=1\cdots I$. Each $y\ind{i}(t)$ is then sampled using TEM$\ind{i}$ which produces spike times $\spikes{i}$.}
    \label{fig: mixed TEM setup}
    \end{figure}

To do so, we build on a result from \cite{adam2020encoding, adam2021asynchrony} on multi-channel time encoding of mixed low-dimensional signals. When $I$ signals observed by $I$ TEMs have a lower dimensional representation and can be written as a linear combination of $J\leq I$ signals, as depicted in Fig.~\ref{fig: mixed TEM setup}, the low dimensionality of the input can be used to reduce the number of spike pairs needed to ensure reconstruction of the input.

To formalize this, let $\vect{y}(t)$ denote an input vector signal composed of $y\ind{i}(t)$, $i=1,\cdots,I$, such that
\begin{assumptions}
\item each $y\ind{i}(t)$ is a $T$-periodic bandlimited signal:
\begin{equation}
y\ind{i}(t) = \sum_{k=-K_0}^{K_0} c_{i,k}(\vect{y}) \exp\left(\mathbf{j} \frac{2\pi k}{T}t\right),
\end{equation}
where the $c_{i,k}(\vect{y})$ are fixed coefficients that are unknown apriori and form a matrix $\vect{C}(\vect{y})$ \label{assume: parametric},

\item each $y\ind{i}(t)$ can be written as a linear combination of $x\ind{j}(t)$'s, $j = 1\cdots J$, where $J\leq I$:
\begin{equation}
\vect{y}(t) = \vect{Ax}(t),
\end{equation}
for $x\ind{j}(t)$'s characterized by a matrix of coefficients $\vect{C}(\vect{x})$ (where the $x\ind{j}(t)$'s and $\vect{C}(\vect{x})$ are unknown apriori) and a known mixing matrix $\vect{A} \in \mathbb{R}^{I\times J}$, and \label{assume: low-dim}

\item each $y\ind{i}(t)$ is sampled using a time encoding machine TEM$\ind{i}$ with known parameters $\kappa\ind{i}$, $\delta\ind{i}$ and $\biasi{i}$, and outputs $\spikes{i}$. \label{assume: time encoded}
\end{assumptions}

Under these assumptions, the input signal $\vect{y}(t)$ can be recovered if the obtained spike times provide a rich enough set of information.
\begin{thm}[Adam, Scholefield and Vetterli 2021]
\label{thm: perfect reconstruction}
Let $\vect{y}(t)$ be an input vector signal be comprised of $I$ signals $y\ind{i}(t), i = 1\cdots I$, satisfying assumptions~\ref{assume: parametric},~\ref{assume: low-dim} and~\ref{assume: time encoded}, with the corresponding coefficients  $c_{j,k}(\vect{x})$ being drawn from a Lipschitz continuous probability distribution. Now assume $\vect{A} \in \mathbb{R}^{I\times J}$ as defined in~\ref{assume: low-dim} has every $J$ rows linearly independent. Then the inputs $y\ind{i}(t), i =1\cdots I$ are exactly determined by the spike times $\spikes{i}, i=1\cdots I$, with probability one if:
\begin{equation}
\label{eq: condition perf rec}
\sum_{i=1}^I \min \left(n_{\mathrm{spikes}^{(i)}}-1, K \right)  > JK.
\end{equation}
	where $K=2K_0+1$.
\end{thm}

This results shows that a signal with low dimensional representation can be reconstructed from its time encoding if the number of \emph{pairs} of spikes scales with the number of parameters in the \emph{low dimensional} space rather than the high dimensional space. In total, one requires $JK\leq IK$ linearly independent constraints where $J$ is the number of signals in the underlying low dimensional representation and $I$ is the number of signals seen by the TEMs.

We will show that we can formulate the problem of time encoding video to fit the framework presented in Assumptions~\ref{assume: parametric}-\ref{assume: time encoded}, allowing us to use Theorem~\ref{thm: perfect reconstruction}.

%% file: figures/TEMCircuit_Tikz.tex
\begin{tikzpicture}[scale = 1]

    \tikzmath{
        \inputstartx = 0;
        \inputstarty = 0.4;
        \inputlen = 1;
    }
    \draw[->] (\inputstartx, \inputstarty) node[anchor = south west] {$y(t)$} -- (\inputlen, \inputstarty);

    \tikzmath{
        \circleradius = 0.25;
        \circlecenterx = \inputstartx+\inputlen+\circleradius;
        \circlecentery = \inputstarty;
    }
    \draw (\circlecenterx, \circlecentery) circle (\circleradius) node {$+$};

    \tikzmath{
        \biaslinelen = 0.5;
        \biaslinex = \circlecenterx;
        \biasliney = \circlecentery -  \circleradius - \biaslinelen;
    }
    \draw[->] (\biaslinex, \biasliney) node[below] {$\bias$} --  (\biaslinex, \biasliney+ \biaslinelen);

    \tikzmath{
        \tointlinex = \circlecenterx + \circleradius;
        \tointliney = \circlecentery;
        \tointlinelen = 0.5;
    }
    \draw[->] (\tointlinex, \tointliney) -- (\tointlinex + \tointlinelen, \tointliney);

    \tikzmath{
        \integboxheight = 1;
        \integboxwidth = 1.5;
        \integboxx = \tointlinex + \tointlinelen;
        \integboxy = \tointliney - \integboxheight/2;
    }
    
    \draw (\integboxx, \integboxy) rectangle (\integboxx+\integboxwidth, \integboxy+\integboxheight) node[midway] {$\frac{1}{\kappa}\int$};

    \tikzmath{
        \tocomplinex = \integboxx + \integboxwidth;
        \tocompliney = \tointliney;
        \tocomplinelen = 1.5;
    }
    \draw[->] (\tocomplinex, \tocompliney) node[anchor = south west] {$z(t)$} -- (\tocomplinex + \tocomplinelen, \tocompliney);
    
    \tikzmath{
        \compboxheight = 1;
        \compboxwidth = 1.5;
        \compboxx = \tocomplinex + \tocomplinelen;
        \compboxy = \tocompliney - \compboxheight/2;
    }
    
    \draw (\compboxx, \compboxy) rectangle (\compboxx+\compboxwidth, \compboxy+\compboxheight) node[midway] {\Large $>$};
    
    \tikzmath{
        \deltalinkx = \compboxx;
        \deltalinky = \compboxy + 0.75\compboxheight;
        \deltalinkwidth = 0.5;
        \deltalinkheight = 0.25;
    }
    \draw[->] (\deltalinkx - \deltalinkwidth, \deltalinky + \deltalinkheight) -- (\deltalinkx- \deltalinkwidth, \deltalinky) -- (\deltalinkx, \deltalinky);
    
    \tikzmath{
        \deltaboxwidth = 0.5;
        \deltaboxheight = 0.5;
        \deltaboxx = \deltalinkx - \deltalinkwidth - \deltaboxwidth/2;
        \deltaboxy = \deltalinky + \deltalinkheight ;
    }
    
    \draw (\deltaboxx, \deltaboxy) rectangle  node {$\delta$} (\deltaboxx + \deltaboxwidth, \deltaboxy+ \deltaboxheight);
    
    \tikzmath{
        \tooutlinex = \compboxx + \compboxwidth;
        \tooutliney = \tocompliney;
        \tooutlinelen = 1.5;
    }
    \draw[->] (\tooutlinex, \tooutliney) node[anchor = south west] {$t_\ell$} -- (\tooutlinex + \tooutlinelen, \tooutliney);
    
    \tikzmath{
        \feedbacklinex = \tooutlinex +0.5;
        \feedbackliney = \tooutliney;
        \feedbacklineheight = 1;
        \feedbacklineendy = \integboxy;
        \feedbacklineendx = \integboxx + \integboxwidth/2;
    }
    \draw[dashed, ->] (\feedbacklinex, \feedbackliney) -- (\feedbacklinex , \feedbackliney- \feedbacklineheight) -- node[below] {\textit{Spike triggered reset}} (\feedbacklineendx , \feedbackliney - \feedbacklineheight) --    (\feedbacklineendx, \feedbacklineendy);
\end{tikzpicture}

%% file: figures/Mixed_TEM.tex
\newcommand{\TEMRectWidth}{1.8}
\newcommand{\TEMRectHeight}{0.9}
\newcommand{\TEMStartCoordX}{1}
\newcommand{\TEMStartCoordY}{1}
\newcommand{\TEMVertSpacing}{0.2}
\newcommand{\dotdotdotsize}{0.025}
\newcommand{\dotdotdotspace}{0.1}
\newcommand{\yToTEMLineLength}{0.5}
\newcommand{\XtoYLineXSpan}{3}
\newcommand{\NumXs}{1.5}
\newcommand{\YHorSpace}{1.5}
\newcommand{\XHorSpace}{1.5}
\newcommand{\XInLineSpan}{1.1}
\newcommand{\TEMOutputLineSpan}{0.6}
\newcommand{\TEMSpikeLineSpan}{2}
\newcommand{\TEMOutSpikeVertSpace}{0.3}
\newcommand{\TEMOutSpikeHorSpace}{0.5}
\newcommand{\SpikeHeight}{1}
\newcommand{\ALabelHeight}{0.2}
\begin{center}
\begin{tikzpicture}[scale = 0.65]
    \small
\foreach \n in {0,1,2,3}
{
    
    \tikzmath{
        int \indn;
        \indn= \n +1;
        \startx = 1;
        \starty = 1 - \n*(\TEMRectHeight+\TEMVertSpacing);
    }
    \ifnum \n = 3
        \tikzmath{
            \starty = 1 - 4*(\TEMRectHeight+\TEMVertSpacing);
        }
    \fi
    \tikzmath{
        \endx = 1 + \TEMRectWidth;
        \endy = \starty + \TEMRectHeight;
        \linestartX = (\startx - \yToTEMLineLength);
        \lineY = (\starty+\endy)/2);
        \TEMLabelX = (\startx+\endx)/2;
    }
    \draw (\startx, \starty) rectangle (\endx, \endy);
    \ifnum \n < 3
        \draw (\TEMLabelX, \lineY) node {TEM$\ind{\indn}$};
    \else
        \draw (\TEMLabelX, \lineY) node {TEM$\ind{I}$};
    \fi
    \draw (\linestartX, \lineY) -- (\startx, \lineY) ;
    \draw (\startx+\TEMRectWidth, \lineY) -- (\startx+\TEMRectWidth+\TEMOutputLineSpan, \lineY);
    \tikzmath{
        \startSpikex = \startx+\TEMRectWidth+\TEMOutputLineSpan + \TEMOutSpikeHorSpace;
        \startSpikey = \lineY - \TEMOutSpikeVertSpace;
    }
    \ifnum \n = 0
        \foreach \loc in {0.1,0.4,0.6,1,1.5, 1.7}
        {
            \draw (\startSpikex + \loc, \startSpikey) -- (\startSpikex + \loc, \startSpikey + \SpikeHeight);
        }
    \fi
    \ifnum \n = 1
        \foreach \loc in {0.2,0.7,1.1, 1.6}
        {
            \draw (\startSpikex + \loc, \startSpikey) -- (\startSpikex + \loc, \startSpikey + \SpikeHeight);
        }
    \fi
    \ifnum \n = 2
        \foreach \loc in {0.15,0.5,0.8,1,1.2,1.7}
        {
            \draw (\startSpikex + \loc, \startSpikey) -- (\startSpikex + \loc, \startSpikey + \SpikeHeight);
        }
    \fi
    \ifnum \n = 3
        \foreach \loc in {0.3,1, 1.6}
        {
            \draw (\startSpikex + \loc, \startSpikey) -- (\startSpikex + \loc, \startSpikey + \SpikeHeight);
        }
    \fi
    
    \ifnum \n <3
    \draw (\linestartX, \lineY) node[anchor = east] {$y\ind{\indn}(t)$};
    \draw (\startSpikex, \startSpikey) -- (\startSpikex + \TEMSpikeLineSpan, \startSpikey) node[anchor = south west] {$\quad \left\lbrace t_\ell\ind{\indn}\right\rbrace_\ell$};
    \else
    \draw (\linestartX, \lineY) node[anchor = east] {$y\ind{I}(t)$};
    \draw (\startSpikex, \startSpikey) -- (\startSpikex + \TEMSpikeLineSpan, \startSpikey) node[anchor = south west] {$\quad \left\lbrace t_\ell\ind{I}\right\rbrace_\ell$};
    \fi
}
    
\foreach \n in {-1,0,1}
    \tikzmath{
        \centerx = 1 + \TEMRectWidth/2;
        \centery = 1 - 3*(\TEMRectHeight+\TEMVertSpacing) + 0.5*\TEMRectHeight +\n*(\dotdotdotspace+\dotdotdotsize);
    }
    \fill (\centerx, \centery) circle (\dotdotdotsize);
    

\tikzmath{
    \MTEMTotalHeight = 4*\TEMRectHeight + 3*\TEMVertSpacing;
}

\foreach \n in {0,1,2}
{
    \tikzmath{
        \lineStartX = 1 - \yToTEMLineLength - \XtoYLineXSpan - \YHorSpace;
    }
        \ifnum \n < 2
        \tikzmath{\lineStartY = 1 - \n*(\TEMRectHeight+\TEMVertSpacing) - \TEMVertSpacing/2;}
        \else
        \tikzmath{\lineStartY = 1 - (\n+1)*(\TEMRectHeight+\TEMVertSpacing) - \TEMVertSpacing/2;}
        \fi

    \foreach \m in {0,1,2,3}
    {
        \tikzmath{\lineEndX = 1 - \yToTEMLineLength -\YHorSpace;}
        \ifnum \m < 3
        \tikzmath{\lineEndY = 1 - \m*(\TEMRectHeight+\TEMVertSpacing) +\TEMRectHeight/2;}
        \else
        \tikzmath{\lineEndY = 1 - (\m+1)*(\TEMRectHeight+\TEMVertSpacing) +\TEMRectHeight/2;}
        \fi
        \tikzmath{
        \aX = 2*\lineStartX/4 +\lineEndX/2 ;
        \aY = 2*\lineStartY/4 +\lineEndY/2;
        }
        
        \tikzmath{
            int \indn;
            \indn= \n +1;
            int \indm;
            \indm= \m +1;
        }
        \ifnum \n =0
            \ifnum \m <3
                \draw[blue, thick]  (\lineEndX-0.3, \lineEndY) -- (\lineStartX+0.3, \lineStartY) node[near start, above, font=\boldmath, sloped] {$a_{\indm,\indn}$};
            \else
                \ifnum \m = 3
                \draw[blue, thick]  (\lineEndX-0.3, \lineEndY) -- (\lineStartX+0.3, \lineStartY) node[near start, above, font=\boldmath, sloped] {$a_{I,\indn}$};
                \fi
            \fi
            
        \else
            \draw (\lineStartX+0.3, \lineStartY) -- (\lineEndX-0.3, \lineEndY);
        \fi
    }
    \tikzmath{
        int \indn;
        \indn= \n +1;
    }
    \ifnum \n < 2
    \draw (\lineStartX-\XHorSpace, \lineStartY) node[anchor=west] {$x\ind{\indn}(t)$};
    \else
    \draw (\lineStartX-\XHorSpace, \lineStartY) node[anchor=west] {$x\ind{J}(t)$};
    \fi
}

\end{tikzpicture}
\end{center}

%% file: sections/6_3DSignals.tex
\section{Time Encoding Video: Theory}
\label{sec: 2d space}

\subsection{Video Model}
To tackle the problem of time encoding video, we model video as a continuous  signal $y(d_1, d_2, t)$ which varies in three dimensions: two spatial dimensions $d_1$ and $d_2$ and one temporal dimension along $t$. Such a signal is then sampled using a collection of time encoding machines.

For the remainder of this paper we assume that $y(d_1, d_2, t)$ is periodic bandlimited in all dimensions. Such a signal can be described by a finite number of parameters and can thus potentially be fully characterized by a finite number of ``measurements''. In more mathematical terms, we assume that the input signal to an event-based camera or to a time encoding camera can be written:
\begin{align}
y(d_1, d_2, t) = & \sum_{k_0=-K_0}^{K_0}\sum_{k_1=-K_1}^{K_1}\sum_{k_2=-K_2}^{K_2} c_{k_0,k_1,k_2}(y). \notag\\
&\exp \left(\mathbf{j}2\pi\left(\frac{tk_0}{T} +\frac{d_1k_1}{D_1} + \frac{d_2k_2}{D_2} \right)\right),
\label{eq: def 2d space}
\end{align}

where the $c_{k_0,k_1, k_2}(y)$'s denote the 3D Fourier series coefficients of $y(d_1, d_2,t)$. Note that we assume that $y(d_1, d_2,t)$ has $(2K_0+1)\times (2K_1+1)\times (2K_2+1)$ of these coefficients with periods $T$ in the time dimension, $D_1$ and $D_2$ in the first and second space dimensions, respectively.

While the periodic bandlimited model choice may seem restrictive at first sight, note that frame-based video has limited frame rate and finite pixel separation, thus inherently assuming that the input is ``smooth enough'' between temporal and spatial samples. Moreover, frame-based video records data over finite amounts of time and limited space, and assuming periodicity in the input is a natural way to deal with finite sampling windows.

We have described the signal model, we now focus on the measurement approach. We assume that the scene is recorded using integrate-and-fire time encoding machines. Each TEM$\ind{i}$ observes a specific direction $\vect{d}\ind{i} = (d_1\ind{i}, d_2\ind{i})$ in $2D$ space, where  $d_1\ind{i}, d_2\ind{i} \in\mathbb{R}$, and fires corresponding spikes $t_\ell\ind{i}$.

Then, the input $y\ind{i}(t)$ observed by TEM$\ind{i}$ is
\begin{equation}
    y\ind{i}(t) = y(d_1\ind{i}, d_2\ind{i}, t)
    \label{eq: def input to tem}
    \end{equation}

For example, the pixels can be made to lie on a uniform grid, as illustrated in Fig.~\ref{fig: Figure 1}. 

\subsection{Uniqueness of Video Time Encoding}
Assuming the input is periodic bandlimited allows our problem to fit within the framework we presented in  Section~\ref{sec: background}, where low-dimensional signals are mixed and time-encoded.

\begin{lemma}
    The signals $y\ind{i}(t)$, as defined in~\eqref{eq: def input to tem}, satisfy assumptions~\ref{assume: parametric}-\ref{assume: time encoded}, with $J=(2K_1+1)(2K_2+1)$.
    \end{lemma}
\begin{proof}
We first define proxy signals $x\ind{k_1, k_2}(t)$, with $k_1 \in \left\lbrace -K_1, \cdots, K_1 \right\rbrace$ and $k_2 \in \left\lbrace -K_2, \cdots, K_2 \right\rbrace$:
\begin{equation}
x\ind{k_1, k_2}(t) = \sum_{k_0=-K_0}^{K_0} c_{k_0,k_1,k_2}(y)\exp \left(\mathbf{j}2\pi\left(\frac{tk_0}{T}\right)\right). \label{eq: x_k definition}
\end{equation}
With this definition, we can write the input to each TEM$\ind{i}$ as
\begin{align}
y\ind{i}(t) = & \sum_{k_1=-K_1}^{K_1}\sum_{k_2=-K_2}^{K_2} x\ind{k_1,k_2}(t) \notag\\
& \exp \left(\mathbf{j}2\pi\left(\frac{d_1\ind{i}k_1}{D_1} + \frac{d_2\ind{i}k_2}{D_2} \right)\right).
\label{eq: 2d y=Ax}
\end{align}

We further define an index $j$ that has a unique correspondance to $k_1$ and $k_2$: $j := (k_1+K_1+1)\times(2K_2+1) + (k_2+K_2+1)$.
The index $j$ takes on values $1,\cdots, J$ where $J = (2K_1+1)(2K_2+1)$, and we write $k_1(j)$ and $k_2(j)$ the unique indices $k_1$ and $k_2$ that define $j$, so that
\begin{equation}
    x\ind{j}(t) = x\ind{k_1(j),\, k_2(j)}(t).
    \end{equation}

We further define a matrix $\vect{A}$ with entries
\begin{equation}
    a_{i,j} = \exp \left(\mathbf{j}2\pi\left(\frac{d_1\ind{i}k_1(j)}{D_1} + \frac{d_2\ind{i}k_2(j)}{D_2} \right)\right).
    \label{eq: entries of A definition}
    \end{equation}

Given these definitions, we see that $y\ind{i}(t)$ can be written:
\begin{equation}
    y\ind{i}(t) = \sum_{j=1}^{(2K_1+1)\times (2K_2+1)} a_{i,j} x\ind{j}(t)
\end{equation}

As a result, and given that the inputs $y\ind{i}(t)$ are periodic bandlimited functions that are input to TEMs, assumptions ~\ref{assume: parametric}-\ref{assume: time encoded} are satisfied, with the entries of mixing matrix $\vect{A}$ defined by the known directions $\vect{d}\ind{i}$ the TEMs or pixels are observing, as described in~\eqref{eq: entries of A definition}.
\end{proof}

A similar result to Theorem~\ref{thm: perfect reconstruction} can thus be established:
\begin{corollary}
    \label{corol: video tem}
    Assume a signal $y(d_1,d_2,t)$ is defined as in~\eqref{eq: def 2d space} with the coefficients  $c_{k_0,k_1,k_2}(y)$  being drawn from a Lipschitz continuous probability distribution. Further assume  $y(d_1,d_2,t)$ is sampled using $I\geq J= (2K_1+1)(2K_2+1)$ TEMs observing directions $\vect{d}\ind{i}$ such that the resulting matrix $\vect{A}$ with entries defined in~\eqref{eq: entries of A definition} has every J rows linearly independent. Then, if each TEM$\ind{i}$ fires $n_{\mathrm{spikes}}\ind{i}$ spikes and
    \begin{equation}
        \label{eq: video suff cond}
    \sum_{i=1}^I \min \left(n_{\mathrm{spikes}^{(i)}}-1, K \right)  > JK,
        \end{equation}
    where we defined $K=(2K_0+1)$, the scene $y(d_1,d_2,t)$ can be perfectly reconstructed from its spike times.
    \end{corollary}

Before we dive into the assumptions for this theorem, namely the condition on the mixing matrix $\vect{A}$, let us understand the implications.
First recall that the reconstruction of any input signal to a TEM can be achieved because \emph{pairs} of consecutive spike times provide linear constraints on the input as explained in~\eqref{eq: sig integral}.
According to the result in Corollary~\ref{corol: video tem}, the number $I$ of TEMs used to encode an input signal $y(d_1,d_2, t)$ does not affect of the number of pairs of spike times needed to ensure a unique characterization of the scene, as long as the matrix $\vect{A}$ has every $J$ rows linearly independent. It is rather the complexity $JK = \prod\limits_{n=0}^{2}(2K_n+1)$ of the scene itself, determined by the number of Fourier series coefficients, which dictates the required number of spike time pairs.


We now tackle the requirement that the mixing matrix $\vect{A}$  have every $J=(2K_1+1)(2K_2+1)$ rows linearly independent. In~\cite{adam2021asynchrony}, we showed that a $(2K_1+1)\times (2K_2+1)$ grid of equally spaced pixels results in a matrix $\vect{A}$ that is full rank and therefore fulfills the requirement.
However, this property cannot be extended to uniform grids with more pixels than in the sufficient gridding case. Fortunately, the requirement on matrix  $\vect{A}$ having every $J$ rows linearly independent is merely a \emph{sufficient} requirement for perfect reconstruction rather than a necessary one. We will show in the upcoming simulations that one can still achieve perfect reconstruction under condition~\eqref{eq: video suff cond} even if this requirement on the mixing matrix is not strictly obeyed.

\subsection{Reconstruction Algorithm}
As previously mentioned, each pair of spike times emitted by each TEM$\ind{i}$ provides a linear constraint on the input to this TEM$\ind{i}$. We define a measurement vector $\vect{b}\ind{i}$ associated with each TEM$\ind{i}$ where
\begin{equation}
\label{eq: def b_i_l}
b_\ell\ind{i} := \int_{t_\ell\ind{i}}^{t_{\ell+1}\ind{i}} y\ind{i}(u) \, du,
\end{equation}
where ${b}_\ell\ind{i}$ is known and as described in~\eqref{eq: sig integral}.
We can show that our problem can be rewritten as a rank-one sensing problem~\cite{pacholska2020matrix} and, with some vectorization operations, the right hand side of~\eqref{eq: def b_i_l} can be rewritten to obtain
\begin{equation}
    \label{eq: lin sys to solve}
    b_\ell\ind{i} = \mathrm{vec}\left(\vect{a}_i \left[{\vect{F}}_\ell\ind{i}\right]^T \right)\mathrm{vec}\left(\vect{C}(\vect{x})\right),
    \end{equation}
where $\vect{a}_i$ denotes a row of matrix $\vect{A}$, $\mathrm{vec}()$ denotes the vectorization operation and
\begin{equation*}
    \left[{\vect{F}}_\ell\ind{i})\right]_k = \int_{t_{\ell}\ind{i}}^{t_{\ell+1}\ind{i}} \exp\left(\mathbf{j} \frac{2\pi (k-K_0-1)}{T}u\right)\, du.
\end{equation*}
Therefore, as recovering the video is equivalent to recovering the coefficients $\vect{C}(\vect{x})$,  and as $\vect{A}$ and $\left[{\vect{F}}_\ell\ind{i}\right]$ are known, one can recover the video by simply inverting the linear system in~\eqref{eq: lin sys to solve}. The system is full rank and uniquely invertible if the conditions of Corollary~\ref{corol: video tem} are met.

\begin{figure*}
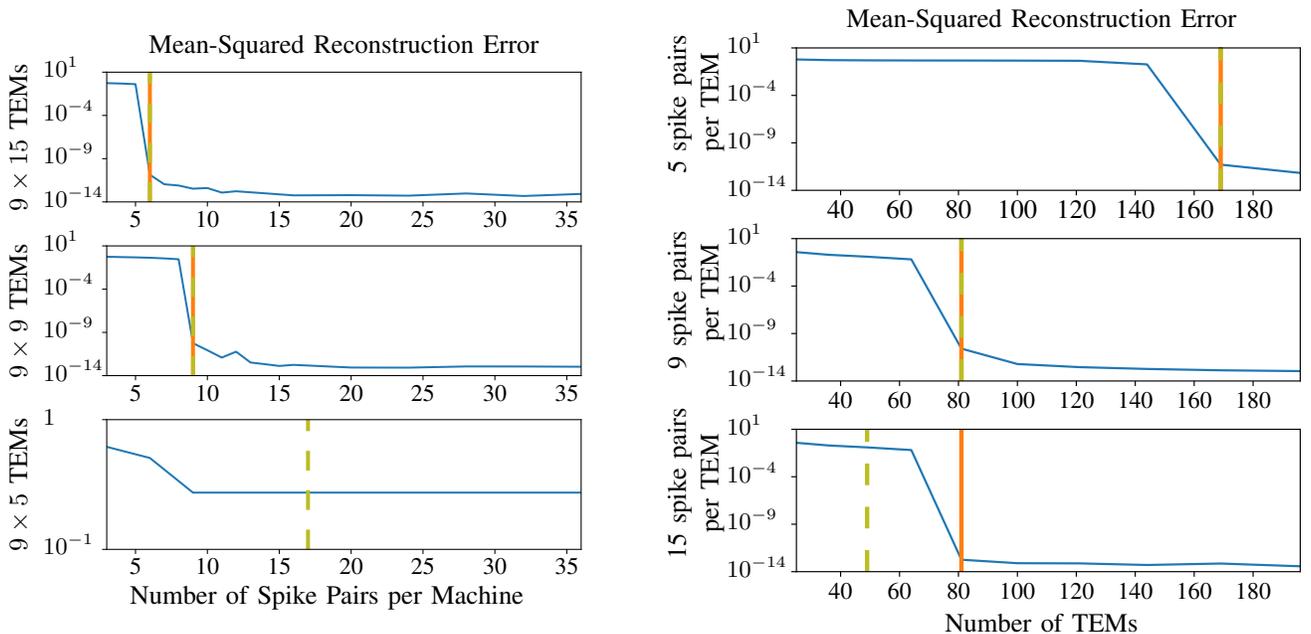

    \centering
    \begin{subfigure}{0.5\textwidth}%
        \centering
    \def\svgwidth{0.82\columnwidth}%
    \subfile{../figures/n_spikes_sweep.tex}%
    \end{subfigure}%
    \begin{subfigure}{0.5\textwidth}%
        \centering
    \def\svgwidth{0.9\columnwidth}%
    \subfile{../figures/n_tems_sweep.tex}%
    \end{subfigure}%
    \caption{ \label{fig: joint spike rate vs n TEMs} Mean-squared reconstruction error with varying number of pixel TEMs and varying number of spike pairs per TEM. The original video has $9\times 9\times 9$ Fourier series coefficients that we wish to recover.
    On the left-hand side, we study the evolution of the error as the number of spikes increases, assuming we time encode the video using grids of uniformly spaced TEMs with sizes that are decreasing from top to bottom, and where the second row assumes the minimal number of TEMs needed to reconstruct the video.
    On the right-hand side, we study the evolution of the error as the number of TEMs increases for numbers of spikes emitted per machine which are increasing from top to bottom, and where the second row assumes the maximal useful spiking rate per TEM.
    For each plot, the vertical orange line marks the point starting from which the condition for Corollary~\ref{corol: video tem} is satisfied. The dashed green line marks the point starting from which we have more constraints than unknowns, without accounting for linear independence, i.e. counting the number of obtained spike pairs rather than estimated the quantity on the left hand side in~\eqref{eq: video suff cond}.
    Note that $N$ spike pairs corresponds to $N+1$ spikes.}%
\end{figure*}

\subsection{Interpretation of Results}
The observation in Corollary~\ref{corol: video tem} provides intuition on how the parameters of the input signal are recovered. Essentially, our result states that if there are more pixels than needed for full spatial resolution (i.e. $I\geq (2K_1+1)(2K_2+1)$), recovery occurs when there are sufficient spike pairs coming from \emph{all} TEMs or pixels, provided that these spike pairs yield non-redundant information.
In fact, our statement is supported by two components of the condition in~\eqref{eq: video suff cond}: (1) the max term ensures that only ``useful'' information is counted from each TEM (given that each TEM has a limited information capacity determined by the complexity of its input) and (2) the summation ensures that the information gathered from the different machines is used collaboratively.

An interesting effect ensues: a machine that spikes too rarely can be compensated for by having other machines spike more often or simply by having \emph{more} machines, and one can thus always improve signal reconstruction by increasing the number of TEMs or pixels used, because the emitted spike times will almost surely be asynchronous.

In some sense, spatial sampling density now has not only an effect on spatial resolution, but on temporal resolution as well, an effect that does not exist in classical frame-based video. This effect occurs because TEMs emit spikes at different times, thus collecting information about their inputs at different times and providing linearly independent constraints on the input. In the frame-based scenario, on the other hand, each frame is captured at \emph{the same time} and increasing the pixel grid size cannot improve temporal resolution.

Therefore, if we have TEM-like receptors or sensors that have a limited spiking rate, spatial and temporal resolution can be regained by adding more sensors that observe new directions. 
In the frame-based scenario, spatial and temporal resolution are independendent of eachother: the former is defined by the pixel grid used in each frame and the latter is defined by the frame rate used to capture the video. On the other hand, employing event-based vision creates an entanglement between spatial and temporal resolution and the latter can also be improved by increasing the size of the pixel grid, as we will show in the upcoming simulations.

\section{Time Encoding Video: Experiments}

In the previous section, we discussed the relationship between spatial sampling density and temporal resolution: in event-based vision, the former influences the latter, whereas in frame-based video, the two quantities are independent.

We would like to show this effect through simulations.
As  mentioned, using  a grid of uniformly spaced pixels and increasing the number of pixels beyond the necessary number $(2K_1+1)(2K_2+1)$ violates the condition on $\vect{A}$ in Corollary~\ref{corol: video tem}. However, we will show that under the same assumptions, the predicitions of the corollary can still be realized.

We time encode a patch of a video recorded with a standard frame-based camera from the Need for Speed dataset~\cite{galoogahi2017need}. The patch is originally 9 pixels high, 9 pixels wide and 9 frames long and we therefore assume that it is periodic bandlimited with $9\times 9 \times 9 $ Fourier series coefficients (where we assume $K_0=K_1=K_2=4$). This assumption allows us to have a continuous model for the video and to perform time encoding of a smooth and continuous input signal.

We sample the smoothly varying patch using different numbers of time encoding machines with different spiking rates and evaluate the result under the different assumptions. Different spiking rates can be achieved by manipulating the threshold of the TEMs: the lower the threshold, the higher the number of spikes emitted over a certain time. Note that emitting more spikes requires more power, so the choice of the threshold always entails a tradeoff between power consumption and reconstruction error.

We place TEMs, for example, at the yellow dots in Fig.~\ref{fig: Figure 1} in a 9$\times$9 grid of time encoding machines.
We will show in our experiments that this is the minimum number of TEMs required to achieve perfect reconstruction.

We will also show how we can use more TEMs in the spatial dimensions to obtain better resolution in the time dimension. This will not necessarily be the case the other way around: more sampling in time does not always provide improved spatial frequency resolution. To achieve this, we run two experiments.

In the first experiment, depicted in the left part of Fig.~\ref{fig: joint spike rate vs n TEMs}, we evaluate the reconstruction performance when the grid of TEMs has more or fewer TEMs. When there are at least as many TEMs as necessary (i.e. a $9\times 9$ or $9\times 15$ grid of TEMs), we see that the reconstruction error indeed decreases sharply when the condition for Corollary~\ref{corol: video tem} is achieved, as indicated by the vertical orange line. When there are fewer TEMs than necessary (a $9\times 5$ grid of TEMs), the spatial density is not sufficient and perfect reconstruction can never be reached as the system will always be underdetermined due to the too few number of sensors.

In the second experiment, depicted in the right part of Fig.~\ref{fig: joint spike rate vs n TEMs}, we evaluate the reconstruction performance given fixed spiking rates of the TEMs. When the spiking rate is at most the maximal rate per TEM (i.e. 9 spikes per TEM allow each machine to perfectly resolve its input), the reconstruction error decreases with the increase of number of TEMs used, when the condition for Corollary~\ref{corol: video tem} is achieved, as indicated by the vertical orange line. When the spiking rate increases beyond the useful rate (in this case, we have 15 spikes per machine), the higher spiking rate only provides redundant information in the noiseless case and the profile of the reconstruction error resembles that in the case of a spiking rate of $9$.

With these experiments, we see how increased spatial density can increase overall reconstruction (including temporal resolution), even if each TEM has a limited spiking rate.

%% file: figures/n_spikes_sweep.tex
\begingroup%
  \makeatletter%
  \providecommand\color[2][]{%
    \errmessage{(Inkscape) Color is used for the text in Inkscape, but the package 'color.sty' is not loaded}%
    \renewcommand\color[2][]{}%
  }%
  \providecommand\transparent[1]{%
    \errmessage{(Inkscape) Transparency is used (non-zero) for the text in Inkscape, but the package 'transparent.sty' is not loaded}%
    \renewcommand\transparent[1]{}%
  }%
  \providecommand\rotatebox[2]{#2}%
  \newcommand*\fsize{\dimexpr\f@size pt\relax}%
  \newcommand*\lineheight[1]{\fontsize{\fsize}{#1\fsize}\selectfont}%
  \ifx\svgwidth\undefined%
    \setlength{\unitlength}{432bp}%
    \ifx\svgscale\undefined%
      \relax%
    \else%
      \setlength{\unitlength}{\unitlength * \real{\svgscale}}%
    \fi%
  \else%
    \setlength{\unitlength}{\svgwidth}%
  \fi%
  \global\let\svgwidth\undefined%
  \global\let\svgscale\undefined%
  \makeatother%
  \begin{picture}(1,1)%
    \lineheight{1}%
    \setlength\tabcolsep{0pt}%
    \put(0,0){\includegraphics[width=\unitlength,page=1]{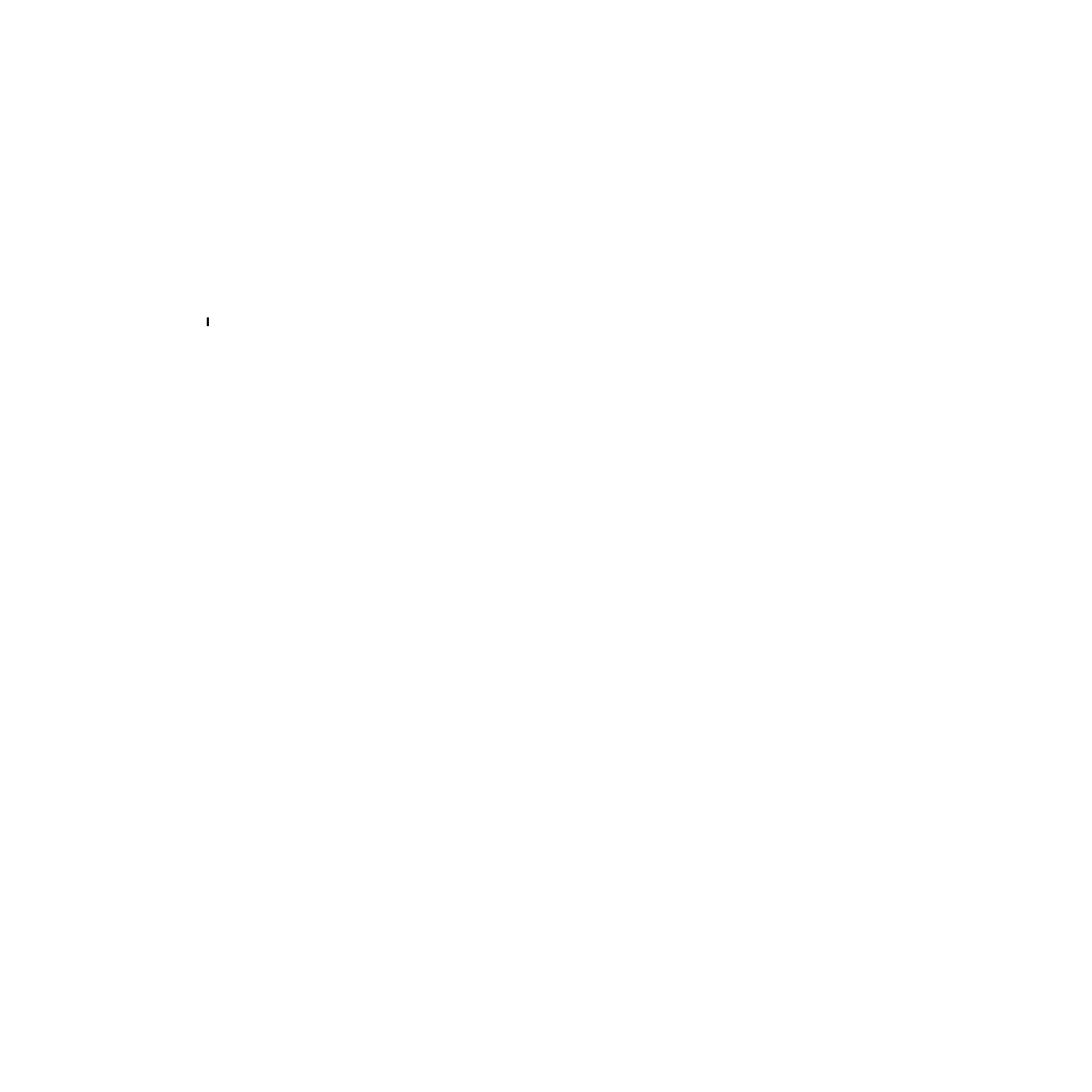}}%
    \put(0.1905326,0.66546039){\makebox(0,0)[t]{\lineheight{1.25}\smash{\begin{tabular}[t]{c}\small 5\end{tabular}}}}%
    \put(0,0){\includegraphics[width=\unitlength,page=2]{n_spikes_sweep_pdf.pdf}}%
    \put(0.31705959,0.66546039){\makebox(0,0)[t]{\lineheight{1.25}\smash{\begin{tabular}[t]{c}\small 10\end{tabular}}}}%
    \put(0,0){\includegraphics[width=\unitlength,page=3]{n_spikes_sweep_pdf.pdf}}%
    \put(0.4435866,0.66546039){\makebox(0,0)[t]{\lineheight{1.25}\smash{\begin{tabular}[t]{c}\small 15\end{tabular}}}}%
    \put(0,0){\includegraphics[width=\unitlength,page=4]{n_spikes_sweep_pdf.pdf}}%
    \put(0.57011361,0.66546039){\makebox(0,0)[t]{\lineheight{1.25}\smash{\begin{tabular}[t]{c}\small 20\end{tabular}}}}%
    \put(0,0){\includegraphics[width=\unitlength,page=5]{n_spikes_sweep_pdf.pdf}}%
    \put(0.69664058,0.66546039){\makebox(0,0)[t]{\lineheight{1.25}\smash{\begin{tabular}[t]{c}\small 25\end{tabular}}}}%
    \put(0,0){\includegraphics[width=\unitlength,page=6]{n_spikes_sweep_pdf.pdf}}%
    \put(0.82316759,0.66546039){\makebox(0,0)[t]{\lineheight{1.25}\smash{\begin{tabular}[t]{c}\small 30\end{tabular}}}}%
    \put(0,0){\includegraphics[width=\unitlength,page=7]{n_spikes_sweep_pdf.pdf}}%
    \put(0.9496946,0.66546039){\makebox(0,0)[t]{\lineheight{1.25}\smash{\begin{tabular}[t]{c}\small 35\end{tabular}}}}%
    \put(0,0){\includegraphics[width=\unitlength,page=8]{n_spikes_sweep_pdf.pdf}}%
    \put(0.02813475,0.7004586){\makebox(0,0)[lt]{\lineheight{1.25}\smash{\begin{tabular}[t]{l}\small $10^{-14}$\end{tabular}}}}%
    \put(0,0){\includegraphics[width=\unitlength,page=9]{n_spikes_sweep_pdf.pdf}}%
    \put(0.02813475,0.7766335){\makebox(0,0)[lt]{\lineheight{1.25}\smash{\begin{tabular}[t]{l}\small $10^{-9}$\end{tabular}}}}%
    \put(0,0){\includegraphics[width=\unitlength,page=10]{n_spikes_sweep_pdf.pdf}}%
    \put(0.02813475,0.85280839){\makebox(0,0)[lt]{\lineheight{1.25}\smash{\begin{tabular}[t]{l}\small $10^{-4}$\end{tabular}}}}%
    \put(0,0){\includegraphics[width=\unitlength,page=11]{n_spikes_sweep_pdf.pdf}}%
    \put(0.02813475,0.92898329){\makebox(0,0)[lt]{\lineheight{1.25}\smash{\begin{tabular}[t]{l}\small $10^{1}$\end{tabular}}}}%
    \put(0,0.82351543){\rotatebox{90}{\makebox(0,0)[t]{\lineheight{1.25}\smash{\begin{tabular}[t]{c}$9\times15$ TEMs\end{tabular}}}}}%
    \put(0,0){\includegraphics[width=\unitlength,page=12]{n_spikes_sweep_pdf.pdf}}%
    \put(0.55746089,0.97166667){\makebox(0,0)[t]{\lineheight{1.25}\smash{\begin{tabular}[t]{c}Mean-Squared Reconstruction Error\end{tabular}}}}%
    \put(0,0){\includegraphics[width=\unitlength,page=13]{n_spikes_sweep_pdf.pdf}}%
    \put(0.1905326,0.35939868){\makebox(0,0)[t]{\lineheight{1.25}\smash{\begin{tabular}[t]{c}\small 5\end{tabular}}}}%
    \put(0,0){\includegraphics[width=\unitlength,page=14]{n_spikes_sweep_pdf.pdf}}%
    \put(0.31705959,0.35939868){\makebox(0,0)[t]{\lineheight{1.25}\smash{\begin{tabular}[t]{c}\small 10\end{tabular}}}}%
    \put(0,0){\includegraphics[width=\unitlength,page=15]{n_spikes_sweep_pdf.pdf}}%
    \put(0.4435866,0.35939868){\makebox(0,0)[t]{\lineheight{1.25}\smash{\begin{tabular}[t]{c}\small 15\end{tabular}}}}%
    \put(0,0){\includegraphics[width=\unitlength,page=16]{n_spikes_sweep_pdf.pdf}}%
    \put(0.57011361,0.35939868){\makebox(0,0)[t]{\lineheight{1.25}\smash{\begin{tabular}[t]{c}\small 20\end{tabular}}}}%
    \put(0,0){\includegraphics[width=\unitlength,page=17]{n_spikes_sweep_pdf.pdf}}%
    \put(0.69664058,0.35939868){\makebox(0,0)[t]{\lineheight{1.25}\smash{\begin{tabular}[t]{c}\small 25\end{tabular}}}}%
    \put(0,0){\includegraphics[width=\unitlength,page=18]{n_spikes_sweep_pdf.pdf}}%
    \put(0.82316759,0.35939868){\makebox(0,0)[t]{\lineheight{1.25}\smash{\begin{tabular}[t]{c}\small 30\end{tabular}}}}%
    \put(0,0){\includegraphics[width=\unitlength,page=19]{n_spikes_sweep_pdf.pdf}}%
    \put(0.9496946,0.35939868){\makebox(0,0)[t]{\lineheight{1.25}\smash{\begin{tabular}[t]{c}\small 35\end{tabular}}}}%
    \put(0,0){\includegraphics[width=\unitlength,page=20]{n_spikes_sweep_pdf.pdf}}%
    \put(0.02813475,0.39439687){\makebox(0,0)[lt]{\lineheight{1.25}\smash{\begin{tabular}[t]{l}\small $10^{-14}$\end{tabular}}}}%
    \put(0,0){\includegraphics[width=\unitlength,page=21]{n_spikes_sweep_pdf.pdf}}%
    \put(0.02813475,0.47057177){\makebox(0,0)[lt]{\lineheight{1.25}\smash{\begin{tabular}[t]{l}\small $10^{-9}$\end{tabular}}}}%
    \put(0,0){\includegraphics[width=\unitlength,page=22]{n_spikes_sweep_pdf.pdf}}%
    \put(0.02813475,0.54674666){\makebox(0,0)[lt]{\lineheight{1.25}\smash{\begin{tabular}[t]{l}\small $10^{-4}$\end{tabular}}}}%
    \put(0,0){\includegraphics[width=\unitlength,page=23]{n_spikes_sweep_pdf.pdf}}%
    \put(0.02813475,0.62292156){\makebox(0,0)[lt]{\lineheight{1.25}\smash{\begin{tabular}[t]{l}\small $10^{1}$\end{tabular}}}}%
    \put(0,0.5174537){\rotatebox{90}{\makebox(0,0)[t]{\lineheight{1.25}\smash{\begin{tabular}[t]{c}$9\times9$ TEMs\end{tabular}}}}}%
    \put(0,0){\includegraphics[width=\unitlength,page=24]{n_spikes_sweep_pdf.pdf}}%
    \put(0.1905326,0.05333697){\makebox(0,0)[t]{\lineheight{1.25}\smash{\begin{tabular}[t]{c}\small 5\end{tabular}}}}%
    \put(0,0){\includegraphics[width=\unitlength,page=25]{n_spikes_sweep_pdf.pdf}}%
    \put(0.31705959,0.05333697){\makebox(0,0)[t]{\lineheight{1.25}\smash{\begin{tabular}[t]{c}\small 10\end{tabular}}}}%
    \put(0,0){\includegraphics[width=\unitlength,page=26]{n_spikes_sweep_pdf.pdf}}%
    \put(0.4435866,0.05333697){\makebox(0,0)[t]{\lineheight{1.25}\smash{\begin{tabular}[t]{c}\small 15\end{tabular}}}}%
    \put(0,0){\includegraphics[width=\unitlength,page=27]{n_spikes_sweep_pdf.pdf}}%
    \put(0.57011361,0.05333697){\makebox(0,0)[t]{\lineheight{1.25}\smash{\begin{tabular}[t]{c}\small 20\end{tabular}}}}%
    \put(0,0){\includegraphics[width=\unitlength,page=28]{n_spikes_sweep_pdf.pdf}}%
    \put(0.69664058,0.05333697){\makebox(0,0)[t]{\lineheight{1.25}\smash{\begin{tabular}[t]{c}\small 25\end{tabular}}}}%
    \put(0,0){\includegraphics[width=\unitlength,page=29]{n_spikes_sweep_pdf.pdf}}%
    \put(0.82316759,0.05333697){\makebox(0,0)[t]{\lineheight{1.25}\smash{\begin{tabular}[t]{c}\small 30\end{tabular}}}}%
    \put(0,0){\includegraphics[width=\unitlength,page=30]{n_spikes_sweep_pdf.pdf}}%
    \put(0.9496946,0.05333697){\makebox(0,0)[t]{\lineheight{1.25}\smash{\begin{tabular}[t]{c}\small 35\end{tabular}}}}%
    \put(0.52746089,0.00167463){\makebox(0,0)[t]{\lineheight{1.25}\smash{\begin{tabular}[t]{c}Number of Spike Pairs per Machine\end{tabular}}}}%
    \put(0,0){\includegraphics[width=\unitlength,page=31]{n_spikes_sweep_pdf.pdf}}%
    \put(0.02813475,0.08833514){\makebox(0,0)[lt]{\lineheight{1.25}\smash{\begin{tabular}[t]{l}\small $10^{-1}$\end{tabular}}}}%
    \put(0,0){\includegraphics[width=\unitlength,page=32]{n_spikes_sweep_pdf.pdf}}%
    \put(0.02813475,0.31685983){\makebox(0,0)[lt]{\lineheight{1.25}\smash{\begin{tabular}[t]{l}\small $1$\end{tabular}}}}%
    \put(0,0.21139197){\rotatebox{90}{\makebox(0,0)[t]{\lineheight{1.25}\smash{\begin{tabular}[t]{c}$9\times5$ TEMs\end{tabular}}}}}%
    \put(0,0){\includegraphics[width=\unitlength,page=33]{n_spikes_sweep_pdf.pdf}}%
  \end{picture}%
\endgroup%

%% file: figures/n_tems_sweep.tex
\begingroup%
  \makeatletter%
  \providecommand\color[2][]{%
    \errmessage{(Inkscape) Color is used for the text in Inkscape, but the package 'color.sty' is not loaded}%
    \renewcommand\color[2][]{}%
  }%
  \providecommand\transparent[1]{%
    \errmessage{(Inkscape) Transparency is used (non-zero) for the text in Inkscape, but the package 'transparent.sty' is not loaded}%
    \renewcommand\transparent[1]{}%
  }%
  \providecommand\rotatebox[2]{#2}%
  \newcommand*\fsize{\dimexpr\f@size pt\relax}%
  \newcommand*\lineheight[1]{\fontsize{\fsize}{#1\fsize}\selectfont}%
  \ifx\svgwidth\undefined%
    \setlength{\unitlength}{432bp}%
    \ifx\svgscale\undefined%
      \relax%
    \else%
      \setlength{\unitlength}{\unitlength * \real{\svgscale}}%
    \fi%
  \else%
    \setlength{\unitlength}{\svgwidth}%
  \fi%
  \global\let\svgwidth\undefined%
  \global\let\svgscale\undefined%
  \makeatother%
  \begin{picture}(1,1)%
    \lineheight{1}%
    \setlength\tabcolsep{0pt}%
    \put(0,0){\includegraphics[width=\unitlength,page=1]{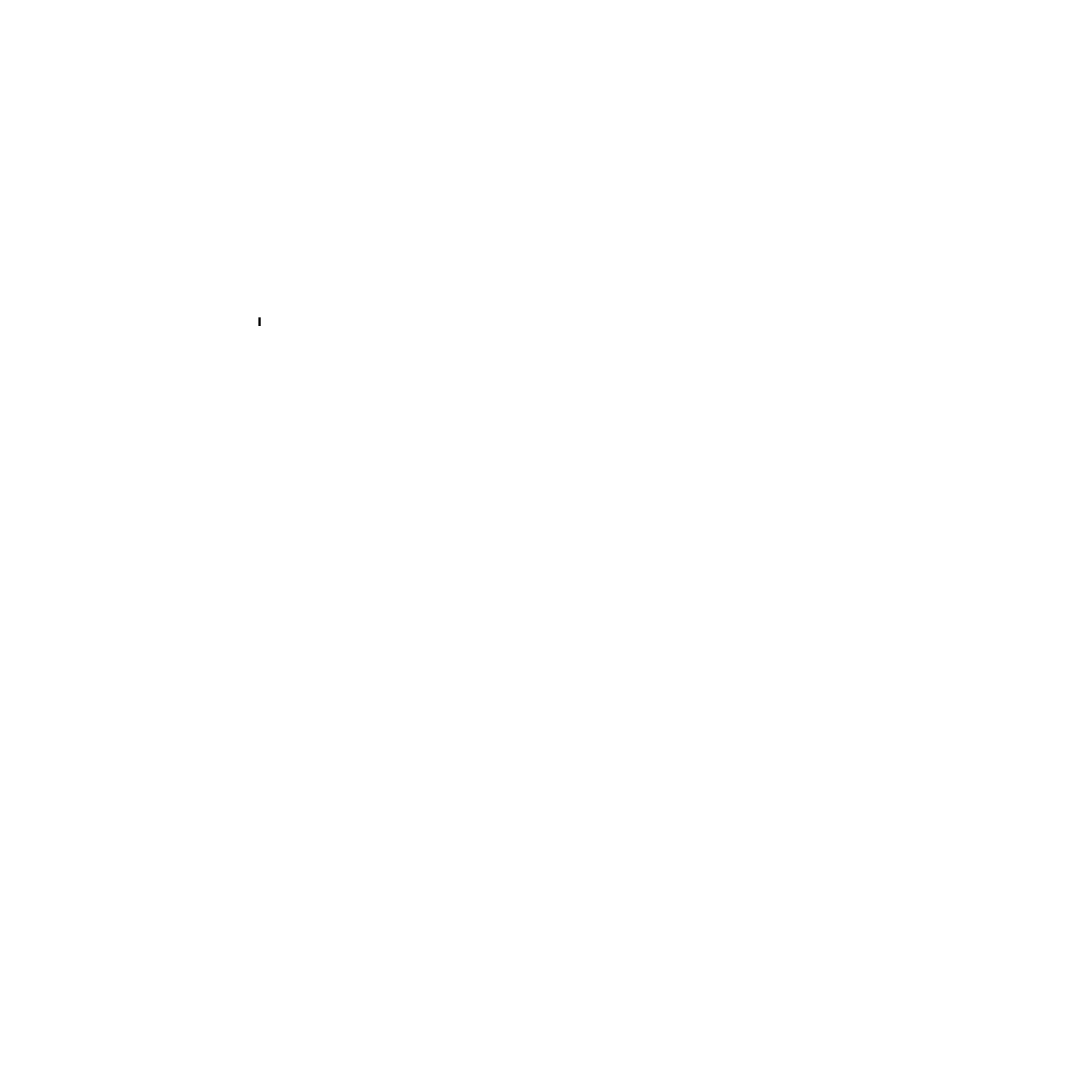}}%
    \put(0,0){\includegraphics[width=\unitlength,page=2]{n_tems_sweep_pdf.pdf}}%
    \put(0,0){\includegraphics[width=\unitlength,page=3]{n_tems_sweep_pdf.pdf}}%
    \put(0,0){\includegraphics[width=\unitlength,page=4]{n_tems_sweep_pdf.pdf}}%
    \put(0,0){\includegraphics[width=\unitlength,page=5]{n_tems_sweep_pdf.pdf}}%
    \put(0,0){\includegraphics[width=\unitlength,page=6]{n_tems_sweep_pdf.pdf}}%
    \put(0,0){\includegraphics[width=\unitlength,page=7]{n_tems_sweep_pdf.pdf}}%
    \put(0,0){\includegraphics[width=\unitlength,page=8]{n_tems_sweep_pdf.pdf}}%
    \put(0,0){\includegraphics[width=\unitlength,page=9]{n_tems_sweep_pdf.pdf}}%
    \put(0.23780584,0.66546039){\makebox(0,0)[t]{\lineheight{1.25}\smash{\begin{tabular}[t]{c}40\end{tabular}}}}%
    \put(0.33231792,0.66546039){\makebox(0,0)[t]{\lineheight{1.25}\smash{\begin{tabular}[t]{c}60\end{tabular}}}}%
    \put(0.42682997,0.66546039){\makebox(0,0)[t]{\lineheight{1.25}\smash{\begin{tabular}[t]{c}80\end{tabular}}}}%
    \put(0.52134207,0.66546039){\makebox(0,0)[t]{\lineheight{1.25}\smash{\begin{tabular}[t]{c}100\end{tabular}}}}%
    \put(0.61585412,0.66546039){\makebox(0,0)[t]{\lineheight{1.25}\smash{\begin{tabular}[t]{c}120\end{tabular}}}}%
    \put(0.71036621,0.66546039){\makebox(0,0)[t]{\lineheight{1.25}\smash{\begin{tabular}[t]{c}140\end{tabular}}}}%
    \put(0.80487831,0.66546039){\makebox(0,0)[t]{\lineheight{1.25}\smash{\begin{tabular}[t]{c}160\end{tabular}}}}%
    \put(0.89939033,0.66546039){\makebox(0,0)[t]{\lineheight{1.25}\smash{\begin{tabular}[t]{c}180\end{tabular}}}}%
    \put(0,0){\includegraphics[width=\unitlength,page=10]{n_tems_sweep_pdf.pdf}}%
    \put(0.05513475,0.7004586){\makebox(0,0)[lt]{\lineheight{1.25}\smash{\begin{tabular}[t]{l}\small $10^{-14}$\end{tabular}}}}%
    \put(0,0){\includegraphics[width=\unitlength,page=11]{n_tems_sweep_pdf.pdf}}%
    \put(0.05513475,0.7766335){\makebox(0,0)[lt]{\lineheight{1.25}\smash{\begin{tabular}[t]{l}\small $10^{-9}$\end{tabular}}}}%
    \put(0,0){\includegraphics[width=\unitlength,page=12]{n_tems_sweep_pdf.pdf}}%
    \put(0.05513475,0.85280839){\makebox(0,0)[lt]{\lineheight{1.25}\smash{\begin{tabular}[t]{l}\small $10^{-4}$\end{tabular}}}}%
    \put(0,0){\includegraphics[width=\unitlength,page=13]{n_tems_sweep_pdf.pdf}}%
    \put(0.05513475,0.92898329){\makebox(0,0)[lt]{\lineheight{1.25}\smash{\begin{tabular}[t]{l}\small $10^{1}$\end{tabular}}}}%
    \put(-0.01,0.73644693){\rotatebox{90}{\makebox(0,0)[lt]{\lineheight{1.25}\smash{\begin{tabular}[t]{l}5 spike pairs\end{tabular}}}}}%
    \put(0.0420614,0.77256238){\rotatebox{90}{\makebox(0,0)[lt]{\lineheight{1.25}\smash{\begin{tabular}[t]{l}per TEM\end{tabular}}}}}%
    \put(0,0){\includegraphics[width=\unitlength,page=14]{n_tems_sweep_pdf.pdf}}%
    \put(0.56002338,0.97166667){\makebox(0,0)[t]{\lineheight{1.25}\smash{\begin{tabular}[t]{c}Mean-Squared Reconstruction Error\end{tabular}}}}%
    \put(0,0){\includegraphics[width=\unitlength,page=15]{n_tems_sweep_pdf.pdf}}%
    \put(0,0){\includegraphics[width=\unitlength,page=16]{n_tems_sweep_pdf.pdf}}%
    \put(0,0){\includegraphics[width=\unitlength,page=17]{n_tems_sweep_pdf.pdf}}%
    \put(0,0){\includegraphics[width=\unitlength,page=18]{n_tems_sweep_pdf.pdf}}%
    \put(0,0){\includegraphics[width=\unitlength,page=19]{n_tems_sweep_pdf.pdf}}%
    \put(0,0){\includegraphics[width=\unitlength,page=20]{n_tems_sweep_pdf.pdf}}%
    \put(0,0){\includegraphics[width=\unitlength,page=21]{n_tems_sweep_pdf.pdf}}%
    \put(0,0){\includegraphics[width=\unitlength,page=22]{n_tems_sweep_pdf.pdf}}%
    \put(0,0){\includegraphics[width=\unitlength,page=23]{n_tems_sweep_pdf.pdf}}%
    \put(0.23780584,0.35939868){\makebox(0,0)[t]{\lineheight{1.25}\smash{\begin{tabular}[t]{c}\small 40\end{tabular}}}}%
    \put(0.33231792,0.35939868){\makebox(0,0)[t]{\lineheight{1.25}\smash{\begin{tabular}[t]{c}\small 60\end{tabular}}}}%
    \put(0.42682997,0.35939868){\makebox(0,0)[t]{\lineheight{1.25}\smash{\begin{tabular}[t]{c}\small 80\end{tabular}}}}%
    \put(0.52134207,0.35939868){\makebox(0,0)[t]{\lineheight{1.25}\smash{\begin{tabular}[t]{c}\small 100\end{tabular}}}}%
    \put(0.61585412,0.35939868){\makebox(0,0)[t]{\lineheight{1.25}\smash{\begin{tabular}[t]{c}\small 120\end{tabular}}}}%
    \put(0.71036621,0.35939868){\makebox(0,0)[t]{\lineheight{1.25}\smash{\begin{tabular}[t]{c}\small 140\end{tabular}}}}%
    \put(0.80487831,0.35939868){\makebox(0,0)[t]{\lineheight{1.25}\smash{\begin{tabular}[t]{c}\small 160\end{tabular}}}}%
    \put(0.89939033,0.35939868){\makebox(0,0)[t]{\lineheight{1.25}\smash{\begin{tabular}[t]{c}\small 180\end{tabular}}}}%
    \put(0,0){\includegraphics[width=\unitlength,page=24]{n_tems_sweep_pdf.pdf}}%
    \put(0.05513475,0.39439687){\makebox(0,0)[lt]{\lineheight{1.25}\smash{\begin{tabular}[t]{l}\small $10^{-14}$\end{tabular}}}}%
    \put(0,0){\includegraphics[width=\unitlength,page=25]{n_tems_sweep_pdf.pdf}}%
    \put(0.05513475,0.47057177){\makebox(0,0)[lt]{\lineheight{1.25}\smash{\begin{tabular}[t]{l}\small $10^{-9}$\end{tabular}}}}%
    \put(0,0){\includegraphics[width=\unitlength,page=26]{n_tems_sweep_pdf.pdf}}%
    \put(0.05513475,0.54674666){\makebox(0,0)[lt]{\lineheight{1.25}\smash{\begin{tabular}[t]{l}\small $10^{-4}$\end{tabular}}}}%
    \put(0,0){\includegraphics[width=\unitlength,page=27]{n_tems_sweep_pdf.pdf}}%
    \put(0.05513475,0.62292156){\makebox(0,0)[lt]{\lineheight{1.25}\smash{\begin{tabular}[t]{l}\small $10^{1}$\end{tabular}}}}%
    \put(-0.01,0.42302119){\rotatebox{90}{\makebox(0,0)[lt]{\lineheight{1.25}\smash{\begin{tabular}[t]{l} 9 spike pairs\end{tabular}}}}}%
    \put(0.0420614,0.46650065){\rotatebox{90}{\makebox(0,0)[lt]{\lineheight{1.25}\smash{\begin{tabular}[t]{l} per TEM\end{tabular}}}}}%
    \put(0,0){\includegraphics[width=\unitlength,page=28]{n_tems_sweep_pdf.pdf}}%
    \put(0,0){\includegraphics[width=\unitlength,page=29]{n_tems_sweep_pdf.pdf}}%
    \put(0,0){\includegraphics[width=\unitlength,page=30]{n_tems_sweep_pdf.pdf}}%
    \put(0,0){\includegraphics[width=\unitlength,page=31]{n_tems_sweep_pdf.pdf}}%
    \put(0,0){\includegraphics[width=\unitlength,page=32]{n_tems_sweep_pdf.pdf}}%
    \put(0,0){\includegraphics[width=\unitlength,page=33]{n_tems_sweep_pdf.pdf}}%
    \put(0,0){\includegraphics[width=\unitlength,page=34]{n_tems_sweep_pdf.pdf}}%
    \put(0,0){\includegraphics[width=\unitlength,page=35]{n_tems_sweep_pdf.pdf}}%
    \put(0.23780584,0.05333697){\makebox(0,0)[t]{\lineheight{1.25}\smash{\begin{tabular}[t]{c}\small 40\end{tabular}}}}%
    \put(0.33231792,0.05333697){\makebox(0,0)[t]{\lineheight{1.25}\smash{\begin{tabular}[t]{c}\small 60\end{tabular}}}}%
    \put(0.42682997,0.05333697){\makebox(0,0)[t]{\lineheight{1.25}\smash{\begin{tabular}[t]{c}\small 80\end{tabular}}}}%
    \put(0.52134207,0.05333697){\makebox(0,0)[t]{\lineheight{1.25}\smash{\begin{tabular}[t]{c}\small 100\end{tabular}}}}%
    \put(0.61585412,0.05333697){\makebox(0,0)[t]{\lineheight{1.25}\smash{\begin{tabular}[t]{c}\small 120\end{tabular}}}}%
    \put(0.71036621,0.05333697){\makebox(0,0)[t]{\lineheight{1.25}\smash{\begin{tabular}[t]{c}\small 140\end{tabular}}}}%
    \put(0.80487831,0.05333697){\makebox(0,0)[t]{\lineheight{1.25}\smash{\begin{tabular}[t]{c}\small 160\end{tabular}}}}%
    \put(0.89939033,0.05333697){\makebox(0,0)[t]{\lineheight{1.25}\smash{\begin{tabular}[t]{c}\small 180\end{tabular}}}}%
    \put(0,0){\includegraphics[width=\unitlength,page=36]{n_tems_sweep_pdf.pdf}}%
    \put(0.56002338,0.00167463){\makebox(0,0)[t]{\lineheight{1.25}\smash{\begin{tabular}[t]{c}Number of TEMs\end{tabular}}}}%
    \put(0,0){\includegraphics[width=\unitlength,page=37]{n_tems_sweep_pdf.pdf}}%
    \put(0.05513475,0.08833514){\makebox(0,0)[lt]{\lineheight{1.25}\smash{\begin{tabular}[t]{l}\small $10^{-14}$\end{tabular}}}}%
    \put(0,0){\includegraphics[width=\unitlength,page=38]{n_tems_sweep_pdf.pdf}}%
    \put(0.05513475,0.16451004){\makebox(0,0)[lt]{\lineheight{1.25}\smash{\begin{tabular}[t]{l}\small $10^{-9}$\end{tabular}}}}%
    \put(0.05513475,0.24068494){\makebox(0,0)[lt]{\lineheight{1.25}\smash{\begin{tabular}[t]{l}\small $10^{-4}$\end{tabular}}}}%
    \put(0.05513475,0.31685983){\makebox(0,0)[lt]{\lineheight{1.25}\smash{\begin{tabular}[t]{l}\small $10^{1}$\end{tabular}}}}%
    \put(-0.01,0.11695947){\rotatebox{90}{\makebox(0,0)[lt]{\lineheight{1.25}\smash{\begin{tabular}[t]{l}15 spike pairs\end{tabular}}}}}%
    \put(0.0420614,0.16043892){\rotatebox{90}{\makebox(0,0)[lt]{\lineheight{1.25}\smash{\begin{tabular}[t]{l}per TEM\end{tabular}}}}}%
  \end{picture}%
\endgroup%

%% file: sections/Conclusion.tex

\section{Conclusion}
\label{sec: Conclusion}

We have shown how to use time encoding to understand event-based video and have consequently demonstrated that event-based vision has an advantage over frame-based vision when it comes to sample complexity.

This advantage arises because event-based cameras emit streams of events from their sensors. As these events are asynchronous across sensors, they provide information about the input that is almost surely linearly independent.
This uncovers a relationship between spatial sampling density and temporal resolution in event-based vision. As sensors emit events at different times, increasing the number of sensors used in event-based video increases both spatial and temporal resolution, without requiring a higher firing rate per sensor.

    We have seen how spikes or events can be sample-efficient way of encoding video and we know that they can be implemented in a power efficient manner~\cite{gallego2019event}. While spikes are more difficult to treat compared to uniform samples, their asynchronous and all-or-none nature provide avenues for improvement over clocked systems~\cite{adam2021time}. We hope to focus future work on finding methods to process spiking or event data in efficient ways.
    \vfill